\documentclass[11pt]{article}
\usepackage{amssymb}
\usepackage{euscript}

\def\a{{\alpha}}
\def\b{{\beta}}

\def\e{{\epsilon}}
\def\bra#1{\langle #1 |}
\def\ket#1{|#1 \rangle}
\def\0{\nonumber}

\def\det{{\rm det}}

\def\log{{\rm log}}

\def\exp{{\rm exp}}

\newcommand\ES{\EuScript{S}}

\newcommand\T{\EuScript{T}}

\newcommand\X{\EuScript{X}}

\newcommand\C{{\cal{C}}}
\newcommand\N{{\cal{N}}}

\newcommand\W{\EuScript{W}}

\newcommand\ee{\end{eqnarray}}      
\newcommand\be{\begin{eqnarray}}
\newcommand\ba{\begin{array}}           
\newcommand\ea{\end{array}}
\newcommand\eeq{\end{equation}}     
\newcommand\beq{\begin{equation}}

\begin{document}
\begin{flushright}
 SISSA/7/05/EP\\ hep-th/0501111
\end{flushright}

\vspace{.1in}
\begin{center}
{\LARGE\bf Fundamental strings in SFT}
\end{center}
\vspace{0.1in}
\begin{center}
L. Bonora\footnote{ bonora@sissa.it},
C.Maccaferri\footnote{maccafer@sissa.it},
R.J.Scherer Santos\footnote{scherer@sissa.it}, D.D.Tolla\footnote{tolla@sissa.it}\\
\vspace{2mm}

{\it International School for Advanced Studies (SISSA/ISAS)\\
Via Beirut 2--4, 34014 Trieste, Italy, and INFN, Sezione di
Trieste}\\
\end{center}
\vspace{0.1in}
\begin{center}
{\bf Abstract}
\end{center}
In this letter we show that vacuum string field theory contains exact solutions
 that can be interpreted as macroscopic fundamental strings. They are formed by
a condensate of infinitely many completely space--localized solutions
(D0-branes).

\section{Introduction}

Vacuum string field theory (VSFT) is a version of Witten's open string field
theory which is conjectured to represent string theory at the tachyon
condensation vacuum \cite{Ras}. Its action is formally the same as the original Witten
theory except that the BRST charge takes a simplified form: it has been argued
that it can be expressed simply in terms of the ghost creation and annihilation
operators. By virtue of this simplification it has been possible to determine
exact classical solutions which have been shown to represent D-branes.
The existence of such solutions confirms the conjecture at the basis of
VSFT. The tachyon condensation vacuum physics can only represent
closed string theory and thus, if VSFT is to represent string theory
at the tachyon condensation vacuum, it should be able to describe closed
string theory in the sense of \cite{Senopenclosed}. The above D-brane solutions, expressed in the open string
language of VSFT, correspond precisely to objects that in closed string
language appear either as boundary state or as solutions of low energy
effective actions.

Recently it has been possible to find in VSFT an exact
time-dependent solution, \cite{BMST}, with the characteristics of
a rolling tachyon \cite{Senroll}. A rolling tachyon describes in
various languages (effective field theory, BCFT, SFT) the decay of
unstable D-branes. It is by now clear that the final product of a
brane decay is formed by massive closed string states. However it
has been shown that, in the presence of a background electric
field also (macroscopic) fundamental strings appear as final
products of a brane decay. Now, since our aim is to describe a brane decay in the framework of VSFT we must show first
of all that such fundamental strings exist as solutions of VSFT.
In this letter we want to present some evidence that such
solutions do exist.

The letter is organized as follows. In the next section we collect some
well--known formulas which are needed in the sequel. In section 3 we show in
a rather informal way how to construct new one-dimensional solutions as
condensate of D0--branes. In section 4 we give a more motivated account of
the same construction by introducing a background $B$ field. In the last
section we provide evidence that the new solutions represent fundamental
strings.

\section{A reminder}

In this section we recall the notation and some useful formulas.
The VSFT action is
\beq
{\cal S}(\Psi)= -  \left(\frac 12 \langle\Psi
|{\cal Q}|\Psi\rangle +
\frac 13 \langle\Psi |\Psi *\Psi\rangle\right)\label{sftaction}
\eeq
where
\beq
{\cal {Q}} =  c_0 + \sum_{n>0} \,(-1)^n \,(c_{2n}+ c_{-2n})\label{calQ}
\eeq
 the ansatz for nonperturbative solutions is in the factorized form
\beq
\Psi= \Psi_m \otimes \Psi_g\label{ans}
\eeq
where $\Psi_g$ and $\Psi_m$ depend purely on ghost and matter
degrees of freedom, respectively. The equation of motion splits into
\be
{\cal Q} \Psi_g & = & - \Psi_g *_g \Psi_g\label{EOMg}\\
\Psi_m & = & \Psi_m *_m \Psi_m\label{EOMm}
\ee
where $*_g$ and $*_m$ refer to the star product involving only the ghost
and matter part.\\
The action for this type of solution is
\beq
{\cal S}(\Psi)= - \frac 1{6 } \langle \Psi_g |{\cal Q}|\Psi_g\rangle
\langle \Psi_m |\Psi_m\rangle \label{actionsliver}
\eeq
$\langle \Psi_m |\Psi_m\rangle$ is the ordinary inner product.
We will  concentrate on the matter part, eq.(\ref{EOMm}), assuming the existence of a universal ghost solution.
The solutions are projectors of the $*_m$ algebra.
The $*_m$ product is defined as follows
\beq
_{123}\!\langle V_3|\Psi_1\rangle_1 |\Psi_2\rangle_2 =_3\!\langle
\Psi_1*_m\Psi_2|,
\label{starm}
\eeq
see \cite{GJ1,Ohta, tope, leclair1} for the definition of the three string vertex $_{123}\!\langle V_3|$.
 In the following we need both translationally
invariant and non-translationally invariant solutions. Although there is a great
variety of such solutions we will stick to those introduced in \cite{RSZ2},
i.e. the sliver and the lump. The former is translationally invariant and
 is defined by
\beq
|\Xi\rangle = \N e^{-\frac 12 a^\dagger\cdot S\cdot a^\dagger}|0\rangle,\quad\quad
a^\dagger\cdot S \cdot a^\dagger = \sum_{n,m=1}^\infty a_n^{\mu\dagger} S_{nm}
 a_m^{\nu\dagger}\eta_{\mu\nu}\label{Xi}
\eeq
where $S= CT$ and
\beq
T= \frac 1{2X} (1+X-\sqrt{(1+3X)(1-X)})\label{sliver}
\eeq
with $X=CV^{11}$.
In proving that this is a solution it is crucial that the matrices
$X=CV^{11},X_+=CV^{12}$ and $X_-=CV^{21}$, are mutually
commuting and commute with $T=CS=SC$, where $C_{nm}= (-1)^n\delta_{nm}$.
The normalization constant $\N$ needs being regularized and is formally
vanishing. It has been shown in other papers how this problem could be dealt with, \cite{BMP1, BMP2}.

The lump solution is engineered to represent a lower dimensional brane
(Dk-branes), therefore it will have ($25-k$)
transverse space directions along which translational
invariance is broken.
Accordingly we split the three string vertex into the tensor product of
the perpendicular part and the parallel part
\be
|V_3\rangle = |V_{3,\perp}\rangle \, \otimes\,|V_{3,_\|}\rangle\label{split}
\ee
The parallel part is the same as in the sliver case while the
perpendicular part is modified as follows.
Following \cite{RSZ2}, we denote by $x^\a,p^\a$, $\a=1,...,k$ the
coordinates and momenta in the transverse directions and introduce the
canonical zero modes oscillators
\be
a_0^{(r)\alpha} = \frac 12 \sqrt b \hat p^{(r)\alpha}
- i\frac {1}{\sqrt b} \hat x^{(r)\alpha},
\quad\quad
a_0^{(r)\alpha\dagger} = \frac 12 \sqrt b \hat p^{(r)\alpha} +
i\frac {1}{\sqrt b}\hat x^{(r)\alpha}, \label{osc}
\ee
where $b$ is a free parameter.
Denoting by $|\Omega_{b}\rangle$ the oscillator vacuum
(\,$a_0^\alpha|\Omega_{b}\rangle=0$\,),
in this new basis the three string vertex is given by
\be
|V_{3,\perp}\rangle'= K\, e^{-E'}|\Omega_b\rangle\label{V3'}
\ee
with
\be
K= \left(\frac {\sqrt{2\pi b^3}}{3(V_{00}+b/2)^2 }\right)^{\frac k2},\quad\quad
E'= \frac 12 \sum_{r,s=1}^3 \sum_{M,N\geq 0} a_M^{(r)\a\dagger}
V_{MN}^{'rs} a_N^{(s)\b\dagger}\eta_{\a\b}\label{E'}
\ee
where $M,N$ denote the couple of indices $\{0,m\}$ and $\{0,n\}$,
respectively.
The coefficients $V_{MN}^{'rs}$ are given in Appendix B of \cite{RSZ2}.
The new Neumann coefficients matrices $V^{'rs}$ satisfy the same relations as
the $V^{rs}$ ones. In particular one can introduce the matrices $X^{'rs}=
C V^{'rs}$, where $C_{NM}=(-1)^N\, \delta_{NM}$, which turn out to commute
with one another. The lump solution $|\Xi'_k\rangle$ has the form (\ref{Xi})
with $S$ along
the parallel directions and $S$ replaced by $S'$ along the perpendicular ones.
In turn $S'=CT'$ and $T'$ has the same form as $T$ eq.(\ref{sliver}) with
$X$ replaced by $X'$. The normalization constant $\N'$ is defined in a way
analogous to $\N$ and the same remarks hold for it. It can be verified that
the ratio of tensions for such solutions is the appropriate one
for $Dk$--branes. Moreover the space profile
of these solutions in the transverse direction is given by a Gaussian
(see \cite{MT,BMS2}). This reinforces the interpretation
of these solutions as branes.

\section{Constructing new solutions}

In this section we would like to show how qualitatively new solutions
to (\ref{EOMm}) can be constructed by accretion of infinite many lumps.
Let us start from a lump solution representing a D0--brane as introduced
in the previous section: it has a Gaussian profile in all space
directions, the form of the string field -- let us denote it $|\Xi_0'\rangle$ --
 will be the same as (\ref{sliver}) with $S$
replaced by $S'$, while the $*$--product will be determined by the primed
three strings vertex (\ref{V3'}). Let us pick one particular space direction,
say the $\a$--th. For simplicity in the following we will drop the
corresponding label from the coordinate $\hat x^\a$, momenta $\hat p^\a$
and oscillators $a^\a$ along this direction. Next we need the same solution
displaced by an amount $s$ in the positive $x$ direction ($x$ being the
eigenvalue of $\hat x$). The appropriate solution has been constructed
by Rastelli, Sen and Zwiebach, \cite{RSZ3}:
\be
\ket{\Xi_0'(s)} =  e^{-is\hat p} \ket{\Xi_0'}\label{Xis}
\ee
It satisfies $\ket{\Xi_0'(s)} * \ket{\Xi_0'(s)} = \ket{\Xi_0'(s)}$.
 Eq.(\ref{Xis}) can be written explicitly as
\be
\ket{\Xi_0'(s)}= \N' e^{-\frac{s^2}{2b}(1-S_{00}')}\,
\exp\left(-\frac {is}{\sqrt{b}}((1-S')\cdot a^\dagger)_0\right)
\, \exp \left(-\frac 12 a^\dagger\cdot S'\cdot a^\dagger\right) \ket{\Omega_b}
\label{Xis'}
\ee
where $((1-S'_{00})\cdot a^\dagger)_0= \sum_{N=0}^\infty ((1-S')_{0N}a_N^\dagger)$ and
$a^\dagger\cdot S'\cdot a^\dagger = \sum_{N,M=0}^\infty a_N^\dagger
S'_{NM} a_M^\dagger$;
$\N'$ is the $\ket{\Xi_0'}$ normalization constant. Moreover one can show
that
\be
\bra{\Xi_0'(s)}\Xi_0'(s)\rangle = \bra{\Xi_0'} \Xi_0'\rangle\label{normXi'}
\ee

The meaning of this solution is better understood if we make its space
profile explicit by contracting it with the coordinate eigenfunction
\be
\ket {\hat x} = \left(\frac 2{\pi b} \right)^{\frac 14} \exp \left(-\frac {x^2}b
-i \frac 2{\sqrt b} a_0^\dagger x+\frac 12 a_0^\dagger a_0^\dagger\right)
\ket{\Omega_b}\label{hatx}
\ee
The result is
\be
\bra {\hat x} \Xi_0'(s)\rangle = \left(\frac 2{\pi b} \right)^{\frac 14}
\frac {\N'}{\sqrt{1+S_{00}'}} e^{-\frac{1-S_{00}'} {1+S_{00}'}\frac {(x-s)^2} {b}}
e^{-\frac {2i}{\sqrt b} \frac{x-s}{1+S_{00}'} S_{0m}'a_m^\dagger}
e^{-\frac 12 a_n^\dagger W_{nm}a_m^\dagger} \ket{0}\label{xprof}
\ee
where $W_{nm}= S'_{nm}- \frac{S'_{n0}S'_{0m}}{1+S'_{00}}$.
It is clear that (\ref{xprof}) represents the same Gaussian profile
as $\ket{\Xi_0'}= \ket{\Xi_0'(0)}$ shifted away from the origin by $s$.

It is important to remark now that two such states $\ket{\Xi_0'(s)}$ and
$\ket{\Xi_0'(s')}$ are $*$--orthogonal and $bpz$--orthogonal provided that
$s\neq s'$. For we have
\be
\ket{\Xi_0'(s)} *\ket{\Xi_0'(s')} = e^{-{\cal C}(s,s')} \ket{\Xi_0'(s,s')}\label{Xiss'}
\ee
where the state $\ket{\Xi_0'(s,s')}$ becomes proportional to $\ket{\Xi_0'(s)}$
when $s=s'$ and needs not be explicitly written down otherwise; while
\be
{\cal C}(s,s')= - \frac 1{2b}\left[ (s^2+ s^{'2}) \left(\frac {T'(1-T')}{1+T'}\right)_{00}
+ss'\left(\frac{(1-T')^2}{1+T'}\right)_{00}\right]\label{Css'}
\ee
The quantity $\left(\frac {T'(1-T')}{1+T'}\right)_{00}$ can be evaluated by using the
basis of eigenvectors of $X'$ and $T'$, \cite{BMST, RSZ4,belov}:
\be
\left(\frac {T'(1-T')}{1+T'}\right)_{00}= 2\int_0^{\infty} dk \, (V_0(k))^2
\frac{t(k) (1-t(k))}{1+t(k)} + \left( V_0^{(\xi)} V_0^{(\xi)} +
V_0^{(\bar\xi)} V_0^{(\bar\xi)} \right)\,\frac{e^{-|\eta|} (1-e^{-|\eta|})}{1+e^{-|\eta|}} \label{00}
\ee
The variable $k$ parametrizes the continuous spectrum and $V_0(k)$ is the relevant
component of the continuous basis. The modulus 1 numbers $\xi$ and $\bar \xi$
parametrize the discrete spectrum and $V_0^{(\xi)}$, $V_0^{(\bar\xi)}$
are the relevant components of the discrete basis (see \cite{belov} for
explicit expressions of the eigenvectors and for the connection between
$\xi$, $\eta$ and $b$). The discrete spectrum part of the RHS of (\ref{00})
is just a number. Let us concentrate on the continuous spectrum contribution.
We have $t(k) = -\exp(-\frac {\pi |k|}2)$. Near $k=0$,
$V_0(k)\sim \frac 12 \sqrt{\frac b{2\pi}}$ and
the integrand $\sim - \frac b{2\pi^2}\frac 1k$,
therefore the integral diverges logarithmically, a singularity we can regularize
with an infrared cutoff $\e$. Taking the signs into account we find that the RHS of
(\ref{00}) goes like $\frac b{2\pi^2}\log\, \e$ as a function of the cutoff.
Similarly one can show that
$\left(\frac{(1-T')^2}{1+T'}\right)_{00}$ goes like $-\frac b{\pi^2}\log\, \e$.
Since for $s\neq s'$, $s^2+s^{'2}> 2 ss'$, we can conclude that
${\cal C}(s,s')\sim
- c\,\log \e$, where $c$ is a positive number. Therefore, when we remove
the cutoff, the factor $e^{-{\cal C}(s,s')}$
vanishes, so that (\ref{Xiss'}) becomes a $*$--orthogonality relation.
Notice that the above logarithmic singularities in the two pieces in the RHS of
(\ref{00}) neatly cancel each other when $s=s'$ and we get the finite number
\be
{\cal C}(s,s) = - \frac {s^2}{2b}(1-S'_{00})\0
\ee
In conclusion we can write
\be
\ket{\Xi_0'(s)} * \ket{\Xi_0'(s')} = \hat \delta(s,s')\ket{\Xi_0'(s)} \label{starortho}
\ee
where $\hat\delta$ is the Kronecker (not the Dirac) delta.

Similarly one can prove that
\be
\bra {\Xi_0'(s')}\Xi_0'(s)\rangle &\!=\!& \frac {\N^2}{\sqrt{\det(1-S^{'2})}}\,
e^{-\frac{s^2+s'^2}{2b}(1-S'_{00})}\, e^{ \frac 1{2b}\left[(s^2+s^{'2})\left(\frac {S'(1-S')}{1+S'}
\right)_{00}+ 2s s' \left(\frac {1-S'}{1+S'}\right)_{00}\right]}\label{bpzss'}
\ee
We can repeat the same argument as above and conclude that
\be
\bra {\Xi_0'(s')}\Xi_0'(s)\rangle =
\hat \delta(s,s')\,\bra {\Xi_0'}\Xi_0'\rangle \label{bpzortho}
\ee

After the above preliminaries, let us consider a sequence $s_1,s_2,\ldots$
of distinct real numbers and the corresponding sequence of displaced D0--branes
$\ket{\Xi_0'(s_n)}$. Due to the property (\ref{starortho}) also the string state
\be
\ket{\Lambda}=\sum_{n=1}^\infty\ket{\Xi_0'(s_n)} \label{Lambda}
\ee
is a solution to (\ref{EOMm}): $\ket{\Lambda} *\ket{\Lambda}= \ket{\Lambda}$.
To figure out what it represents let us study its space profile.
To this end we must sum all the profiles like (\ref{xprof}) and then proceed
to a numerical evaluation. In order to get a one dimensional object, we render the sequence
$s_1,s_2,\ldots$  dense, say, in the positive $x$--axis so that we can
replace the summation with an integral. The relevant integral is
\be
\int_0^\infty ds \,\exp\left[- \a (x-s)^2 - i\b (x-s) \right] =
\frac {\sqrt{\pi}}{2\sqrt{\a}} \left( e^{-\frac {\b^2}{4 \a}}
\left(1+{\rm Erf}\left(
\frac {i\b}{2\sqrt{\a}}+\sqrt{\a} x\right)\right)\right)\label{erf}
\ee
where ${\rm Erf}$ is the error function and
\be
\a= \frac 1b \frac {1-S'_{00}}{1+S'_{00}}, \quad\quad \b= \frac 2{\sqrt{b}}
\frac {S'_{0m} a_m^\dagger}{1+S_{00}'}\0
\ee
Of course (\ref{erf}) is a purely formal expression, but it becomes
meaningful in the $\a'\to 0$ limit. As usual, \cite{MT}, we parametrize this
limit with a dimensionless parameter $\e$ and take $\e\to 0$. Using
the results of \cite{BMST,MT}, one can see that $\a \sim 1/\e$, $\b\sim
1/\sqrt{\e}$, so that $\b/\sqrt{\a}$ tends to a finite limit. Therefore, in this
limit, we can disregard the first addend in the argument of ${\rm Erf}$.
Then, up to normalization, the space profile of $\ket{\Lambda}$
is determined by
\be
\frac 12 \left(1 + {\rm Erf}(\sqrt{\a} x)\right)\label{spprof}
\ee
In the limit $\e\to 0$ this factor tends to a step function valued 1
in  the positive real $x$--axis and 0 in the negative one. Of course a similar
result can be obtained numerically to any degree of accuracy by using
a dense enough discrete $\{ s_n\}$ sequence.

Another way of getting the same result is to use the recipe of \cite{MT}
first on (\ref{xprof}). In this way the middle exponential disappears,
while the first exponential is regularized by hand (remember that
$S_{00}'\to -1$ as $\e\to 0$), so we replace $S_{00}'$ by a parameter
${\mathfrak s}$ and keep it $\neq -1$. Now it is easy to sum over $s_n$.
Again we replace the summation by an integration and see immediately that
the space profile becomes the same as (\ref{spprof}).

Let us stress that the derivation of the space profile in the low energy
regime we have given above is far from rigorous. This is due to the very
singular nature of the lump in this limit, first pointed out by \cite{MT}.
A more satisfactory derivation will be provided in the next section
after introducing a background $B$ field.

In summary, the state $\ket{\Lambda}$ is a solution to (\ref{EOMm}),
which represents, in the low energy limit, a one--dimensional object
with a constant profile that extends from the origin to infinity in the
$x$--direction. Actually the initial point could be any finite point
of the $x$--axis, and it is not hard to figure out how to construct a configuration
that extend from $-\infty$ to $+\infty$. How should we interpret these
condensate of D0--branes? In the absence of supersymmetry
it is not easy to distinguish between D--strings and F--strings (see, for
instance, \cite{schwarz} for a comparison), however in the
last section we will provide some evidence that the one--dimensional
solutions of the type $\ket{\Lambda}$ can be interpreted as  fundamental
strings. This kind of objects are very well--known in string theory
as classical solutions, \cite{CM,G,HLW,DM,MS}, see also \cite{DGHR,HarStro,Dab}.
For the time being let us notice that, due to (\ref{bpzortho}),
\be
\bra{\Lambda} \Lambda\rangle = \sum_{n,m=1}^\infty
\bra{\Xi_0'(s_n)}\Xi_0'(s_m)\rangle = \sum_{n=1}^\infty\,  \bra{\Xi_0'}\Xi_0'\rangle
\ee
It follows that the energy of the solution is infinite. Such an (unnormalized) infinity
is a typical property of fundamental string solutions, see \cite{CM}.

\section{An improved construction}

In this section we would like to justify some of the passages utilized
in discussing the space profile of the fundamental string solution in
section 3. The problems in section 3 are linked to the well--known
singularity of the lump space profile, \cite{MT}, which arises in the
low energy limit ($\e\to 0$) and renders some of the manipulations rather
slippery. The origin of this singularity is the denominator $1+S'_{00}$
that appears in many exponentials. Since, when $\e\to 0$, $S_{00}'\to -1$
the exponentials are ill--defined because the series expansions in $1/\e$
are. The best way to regularize them is to
introduce a constant background $B$--field, \cite{W2,Sch,sugino}.
The relevant formulas can be found in \cite{BMS1}. For the purpose
of this paper we introduce a $B$ field along two space directions, say $x$ and
$y$ (our aim is to regularize the solution in the $x$ direction,
but, of course, there is no way to avoid involving in the process
another space direction).

Let us use the notation $x^\a$ with $\a=1,2$ to denote $x,y$ and let us
denote
\be\label{osmetric}
G_{\a\b} = \Delta\delta_{\a\b},\quad\quad \Delta = 1 + (2\pi B)^2
\ee
the open string metric. As is well--known, as far as lump solutions
are concerned, there is an isomorphism of formulas with the ordinary case
by which $X', S',T'$ are replaced, respectively, by $\X, \ES,\T$,
which explicitly depend on $B$. One should never forget that the latter
matrices involve two space directions. We will denote by $\ket{\hat \Xi_0}$
the D0--brane solution in the presence of the $B$ field.

Without writing down all the details, let us see the significant
changes. Let us replace formula (\ref{Xis}) by
\be
\ket{\hat\Xi_0(\{s^\a\})}= e^{-is^\a \hat p_\a}\ket{\hat \Xi_0}\label{hatXis}
\ee
It satisfies $\ket{\hat\Xi_0(s)}*\ket{\hat\Xi_0(s)}=\ket{\hat\Xi_0(s)}$
and $\bra{\hat\Xi_0(s)}{\hat\Xi_0(s)}\rangle =
\bra{\hat\Xi_0}{\hat\Xi_0}\rangle$.
Instead of (\ref{hatx}) we have
\be
\ket {\{\hat x^\a\}} = \left(\frac {2\Delta}{\pi b} \right)^{\frac 12} \exp
\left[\left(-\frac {x^\a x^\b}b
-i \frac 2{\sqrt b} a_0^{\a\dagger}x^\b +\frac 12 a_0^{\a\dagger}
a_0^{\b\dagger}\right)G_{\a\b}\right]
\ket{\Omega_b}\label{hatxy}
\ee
Next we have
\be
\bra {\{\hat x^\a\}} \hat\Xi_0(s)\rangle &\!=\!&
 \left(\frac {2\Delta}{\pi b} \right)^{\frac 12}
\frac {\hat\N}{\sqrt{\det(1+\ES_{00})}} \exp \left[-\frac 1b (x^\a-s^\a)
\left(\frac{1-\ES_{00}} {1+\ES_{00}}\right)_{\a\b}(x^\b-s^\b)\right. \0\\
&&-\left.\frac {2i}{\sqrt b} (x^\a-s^\a)(1+\ES_{00})_{\a\b}
\ES_{0m}{}^\b{}_\gamma\, a_m^{\gamma\dagger}\right]
\exp \left[-\frac 12 a_n^{\a\dagger} \W_{nm,\a\b}a_m^{\b\dagger}\right] \ket{0}
\label{xyprof}
\ee
where $\det(1+\ES_{00})$ means the determinant of the 2x2 matrix
$(1+\ES_{00})_{\a\b}$ and
\be
\W_{nm,\a\b} = \ES_{nm,\a\b} - \ES_{n0,\a}{}^\gamma \left(\frac
1{1+\ES_{00}}\right)_{nm,\gamma\delta}\ES_{0m}{}^\delta{}_\b\label{W}
\ee
The state we start from, i.e. $\ket{\hat\Xi_0(s)}$, and the relevant
space profile, are obtained by setting $s^1=s$ and $s^2=0$ in the previous
formulas.

Next we have an analog of (\ref{Xiss'}) with $\C(s,s')$ replaced by
\be
\hat \C(s,s') = -\frac 1{2b} (s^2+s^{'2}) \left(\frac
{\T(1-\T)}{1+\T}\right)_{00,11} - \frac{ss'}{2b} \left(\frac{(1-\T)^2}{1+\T}
\right)_{00,11}\label{Chat}
\ee
Proceeding in the same way as in section 3 we can prove the analog of
eq.(\ref{starortho}). By using the spectral representation worked out
in \cite{MST} one can show that $\hat \C$ picks up a logarithmic singularity
unless $s=s'$. In a similar way one can prove the analog of (\ref{bpzortho}).

Now let us discuss the properties of
\be
\ket{\hat \Lambda}= \sum_{n=1}^\infty \ket{\hat \Xi_0(s_n)}\0
\ee
in the low energy limit. We refer to (\ref{xyprof}) with $s^1=s$ and $s^2=0$.
The fundamental difference between this formula and (\ref{xprof}) is that in
the low energy limit $\ES_{00,\a\b}$ becomes diagonal and takes on a value
different from --1. More precisely
\be
\ES_{00,\a\b} \to \frac {2|a|-1}{2|a|+1} G_{\a\b},\quad\quad
a= -\frac {\pi^2}{V_{00} +\frac b2} B\0
\ee
see \cite{BMS2}. Therefore the $1+\ES_{00}$ denominators in (\ref{xyprof})
are not dangerous any more. Similarly one can prove that in the same limit
$\ES_{0n}\to 0$. Moreover the $\e$--expansions about these values
are well--defined. Therefore the space profile we are interested in is
\be
\sim \exp[-\frac \mu{b} (x+s)^2 - \frac \mu{b} (y)^2] \,
\exp[-\frac 12 a_n^{\a\dagger} \ES_{nm,\a\b}a_m^{\b\dagger} ]
\ket {0}\label{spaceprof}
\ee
with a finite normalization factor and $\mu=\frac {2|a|-1}{2|a|+1} \Delta$.
Now one can safely integrate $s$ and obtain the result illustrated
in section 3.
This also sheds light on how the resulting state couples to the $B_{\mu\nu}$ field.
Indeed  the length of this one dimensional objects is measured with
the open string metric (\ref{osmetric}), in other words the $B$--field couples to the string by ``stretching'' it.

\section{Fundamental strings}

In this section we would like to discuss the properties of the
$\Lambda$ solutions we found in the previous sections. In order to
justify the claim we made that they represent fundamental strings,
in the sequel we show that they are still solutions if we attach them to
a D--brane. To this end let us pick $\ket{\Lambda}$ as given by (\ref{Lambda})
with $s_n>0$ for all $n$'s. Now let us consider a D24--brane with the only
transverse direction coinciding with the $x$--axis and centered at $x=0$.
The corresponding lump solution has been introduced at the end of section 2
(case $k=24$). Let us call it $\ket{\Xi'_{24}}$. Due to the particular
configuration chosen, it is easy to prove that $\ket{\Xi'_{24}} + \ket{\Lambda}$
is still a solution to (\ref{EOMm}). This is due to the fact that
$\ket{\Xi'_{24}}$ is $*$-orthogonal to the states $\ket{\Xi_0'(s_n)}$ for all $n$'s.
To be even more explicit we can study the space profile of
$\ket{\Xi'_{24}} + \ket{\Lambda}$, assuming the sequence $s_n$ to become
dense in the positive $x$--axis. Using the previous results it is not hard
to see that the overall configuration is a Gaussian centered at $x=0$ in the
$x$ direction (the D24-brane) with an infinite prong attached to it and
extending along the positive $x$--axis. The latter has a
Gaussian profile in all space directions except $x$.

We remark that the condition $s_n>0$ for all $n$'s is important because
$\ket{\Xi'_{24}} + \ket{\Lambda}$ is not anymore a projector
if the $\{s_n\}$ sequence contains 0, since $\ket{\Xi_0'(0)}$ is not $*$--orthogonal
to $\ket{\Xi'_{24}}$. This remark tells us that it not possible to have
solutions representing configurations in which the string crosses the brane
by a finite amount: the string has to stop at the brane.

This is to be contrasted with the configuration obtained by replacing
$ \ket{\Lambda}$ in $\ket{\Xi'_{24}} + \ket{\Lambda}$ with a D1--brane
along the $x$ axis, that is with $\ket{\Xi'_{1}}$. The state we get is
definitely not a solution to the (\ref{EOMm}). This of course
reinforces the interpretation of the $\ket{\Lambda}$ solution as a
fundamental string.

Needless to say it is trivial to generalize the solution of the type
$\ket{\Xi'_{24}} + \ket{\Lambda}$ to lower dimensional branes.

It is worth pointing out that it is also possible to construct string solutions
of finite length. It is enough to choose the sequence $\{s_n\}$ to lie
between two fixed values, say $a$ and $b$ in the $x$--axis, and then `condense'
the sequence between these two points. In the low energy limit the resulting
solution shows precisely a flat profile for $a<x<b$ and a vanishing profile
outside this interval (and of course a Gaussian profile along the other
space direction). This solution is fit to represent a string stretched
between two D--branes located at $x=a$ and $x=b$.

An important property for fundamental strings is the exchange property.
Let us see if it holds for our solutions in a simple example.
We consider first an extension of the solution (\ref{Lambda}) made of two
pieces at right angles. Let us pick two space directions, $x$ and $y$.
We will denote by $\{s_n^x\}$ and $\{s_n^y\}$ a sequence of points
along the positive $x$ and $y$--axis. The string state
\be
\ket{\Lambda^{\pm\pm}} = \ket{\Xi'_0} + \sum_{n=1}^\infty \ket{\Xi'_0(\pm s^x_n)}+
\sum_{n=1}^\infty \ket{\Xi'_0(\pm s^y_n)}\label{lambdaxy}
\ee
is a solution to (\ref{EOMm}). The $\pm\pm$ label refers to the positive
(negative) $x$ and $y$--axis. This state represents an infinite string
stretched along
the positive (negative) $x$ and $y$--axis including the origin.
Now let us construct the string state
\be
\ket{\Xi'_0} + \sum_{n=1}^\infty \ket{\Xi'_0(s^x_n)}+
\sum_{n=1}^\infty \ket{\Xi'_0(-s^x_n)}+
\sum_{n=1}^\infty \ket{\Xi'_0(s^y_n)}+ \sum_{n=1}^\infty \ket{\Xi'_0(-s^y_n)}
\label{cross}
\ee
This is still a solution to (\ref{EOMm}) and can be interpreted in two
ways: either as $\ket{\Lambda^{++}}+ \ket{\Lambda^{--}}$ or
as  $\ket{\Lambda^{+-}}+ \ket{\Lambda^{-+}}$, up to addition to both of
$\ket{\Xi'_0}$ (a bit removed from the origin). This addition costs
the same amount of energy in the two cases,
an amount that vanishes in the continuous limit. Therefore the solution
(\ref{cross}) represents precisely the exchange property of fundamental strings.

So far we have considered only straight one--dimensional solutions
(in terms of space profiles), or at most solutions represented by straight lines
at right angles. However
this is an unnecessary limitation. It is easy to generalize our construction
to any curve in space. For instance, let us consider two directions in space
and let us denote them again $x$ and $y$ ($\hat p^x$ and $\hat p^y$ being the relevant
momentum operators). Let us construct the state
\be
\ket{\Xi_0'(s^x,s^y)} =  e^{-is^x \hat p^x} e^{-is^y\hat p^y} \ket{\Xi_0'}\label{Xisxsy}
\ee
It is evident that this represents a space--localized solution displaced from the origin
by $s^x$ in the positive $x$ direction and $s^y$ in the positive $y$ direction.
Using a suitable sequence $\{s_n^x\}$ and $\{s_n^y\}$, and rendering it dense, we can
construct any curve in the $x-y$ plane, and, as a consequence, write down a solution
to the equation of motion corresponding to this curve. The generalization to
other space dimensions is straightforward.

We would like to remark that, by generalizing the above construction,
one can also construct higher dimensional objects.
For instance one could repeat the accretion construction by adding parallel
D1--branes (that extend,  say, in the $y$ direction) along the $x$--axis.
In this way we end up with a membrane-like configuration (with a flat
profile in the $x,y$--plane), and continue in the same tune with
higher dimensional configurations.

All the solutions we have considered so far are unstable. However the
fundamental string solutions are endowed with a particular property. Since they
end on a D--brane, their endpoints couple to the electromagnetic field
on the brane, \cite{CM,Stro,Witold}, and carry the corresponding charge.
When the D--brane decays there is nothing that prevents the (fundamental)
strings attached to it from decaying themselves. However in the presence of a
background $E$--field, the latter are excited by the coupling with
the $E$--field and persist (or, at least, persist longer
than the other unstable objects).
This phenomenon is described in \cite{MS} in effective field theory and
BCFT language (see also \cite{MST}).

\begin{center}
{\bf Acknowledgments}
\end{center}
 This research was supported by the Italian MIUR
under the program ``Teoria dei Campi, Superstringhe e Gravit\`a''
and by CAPES-Brasil as far as R.J.S.S. is concerned.


\end{document}